\documentclass[12pt]{article}
\usepackage{graphicx}
\usepackage{amssymb,amsmath,amsfonts,amsthm}
\usepackage{epstopdf}
\usepackage[normalem]{ulem}
\usepackage{comment}
\usepackage{hyperref}

\hypersetup{
   colorlinks=true,         % false: boxed links; true: colored links, false is default
   linkcolor=blue,          % color of internal links, red is default
   citecolor=red,        % color of links to bibliography
   urlcolor=Violet            % color of external links, cyan is default
}

\usepackage[usenames,dvipsnames]{color}

\begin{document}

\title{\Large \bf{Axial anomaly and vector meson dominance model} }

\author{Yaroslav~Klopot,$^1$ \footnote{On leave from  Bogolyubov Institute for Theoretical Physics, 03680, Kiev, Ukraine}~~ 
Armen~Oganesian,$^{1,2}$~~
Oleg~Teryaev,$^1$
\footnote{Electronic addresses: \href{mailto:klopot@theor.jinr.ru}{klopot@theor.jinr.ru}, \href{mailto:armen@itep.ru,}{armen@itep.ru,} \href{mailto:teryaev@theor.jinr.ru}{teryaev@theor.jinr.ru}.}
 \vspace{12pt} \\
\it \small $^1$ Bogoliubov Laboratory of Theoretical Physics,\\
\it \small Joint Institute for Nuclear Research, 141980, Dubna, Russia,\\
\it \small  $^2$Institute of Theoretical and Experimental Physics, 117218,  Moscow, Russia.}

\date{}
\maketitle
%\vfill

\begin{abstract}

The dispersive representation of axial anomaly leads to the anomaly sum rules (ASRs), exact nonperturbative relations in QCD. The analytical continuation of the ASRs to the time-like region is performed. The transition form factors of $\pi^0$, $\eta$ and $\eta'$ mesons in this region are calculated. A good agreement with the available experimental data is found. Based on the ASRs, we have provided the foundations for the vector meson dominance model in these processes.\end{abstract}

%\flushbottom \flushleft \large September 2013
\newpage

%\tableofcontents
%\newpage

\section{Introduction}

An important feature of the quantum field theory is the presence of the axial anomaly \cite{Bell:1969ts}. In particular, the two-photon decay of the pion, $\pi^0 \to \gamma\gamma$, is known to be primarily controlled by the axial anomaly, providing quite an exceptional example of the low-energy process completely determined from the fundamental theory.

It is much less known, that the axial anomaly reveals itself also in the processes, which involve virtual photons. In particular, the photon-meson transitions $\gamma\gamma^* \to M$ (where $M$ is a pseudoscalar meson) can be studied by means of the anomaly sum rules (ASRs), which follow from the dispersive representation of the axial anomaly \cite{Dolgov:1971ri,Horejsi:1985qu,Horejsi:1994aj}.  
The ASR approach was applied to study the $\pi^0$, $\eta$ and $\eta'$ transition form factors (TFFs)  in the space-like momentum transfer region ($q^2<0$)  \cite{Klopot:2010ke,Klopot:2011qq,Klopot:2011ai,Klopot:2012hd,Melikhov:2012qp}.

%These sum rules allow to obtain the expressions for the transition form factors (TFFs) at  arbitrary $q^2$ in a model-independent way without relying on the QCD factorization hypothesis. 

The pseudoscalar meson TFFs provide an important information about the QCD dynamics, allowing to test our understanding of the low-energy QCD properties as well as perturbative QCD (pQCD) predictions. Recently, the topic of pseudoscalar meson TFFs have gained a significant interest because of unexpectedly large values of the pion TFF at $|q^2|>10$ GeV$^2$, measured by BABAR Collaboration \cite{Aubert:2009mc}. These data show an excess over the pQCD predicted limit \cite{Lepage:1980fj}, based on collinear factorization and are hard to explain within the QCD \cite{Stefanis:2012yw,Agaev:2010aq}. Although the later data of BELLE Collaboration \cite{Uehara:2012ag} are quite consistent with the conventional theoretical approach, the controversy remains \cite{Denig:2013gda}. The expected high-precision data from BES-III \cite{Asner:2008nq,Unverzagt:2012zz}, KLOE-2 \cite{Babusci:2011bg} (in the space-like region, $q^2<0$) and CLAS \cite{Amaryan:2013osa,Amaryan:2013eja} (in the time-like region, $q^2>0$) as well as further theoretical investigations (especially those valid in both regions) will give us a more complete understanding of the meson TFFs.

In this work we perform the analytical continuation of the ASRs to the time-like region and also study in detail the small $q^2$ region, unreachable by the conventional perturbative QCD approach. 

%The recent advance in the experimental study of the TFFs in the region of small $q^2$  allows to test the theoretical understanding of the strong interactions in this region. %The upcoming data from BES-III, CLAS   

We also found that the ASR in the time-like region leads to the pole, which is close to the $\rho$ meson mass in the pion TFF. This indicates that the axial anomaly leads to some relation between axial and vector channels. 

Moreover, the ASRs give a theoretical foundation for 
the vector meson dominance (VMD) model, which describes the photon-hadron interaction via the transition of the photon (real or virtual) to the intermediate virtual neutral vector meson.

\section{Axial anomaly: from space-like to time-like region}

 The vector-vector-axial triangle graph amplitude, where the axial anomaly occurs, contains an axial current $J_{\alpha 5}$ and two electromagnetic currents $J_{\mu}=\sum\limits_{i=u,d,s} e_i \bar{q_i}\gamma_\mu q_i$ ($e_i$ are quark charges in the units of the absolute value of electron charge),

\begin{equation} \label{VVA} 
T_{\alpha \mu\nu}(k,q)=\int
d^4 x d^4 y e^{(ikx+iqy)} \langle 0|T\{ J_{\alpha 5}(0) J_\mu (x)
J_\nu(y) \}|0\rangle, 
\end{equation}
where $k$ and $q$ are the photons' momenta. In what follows, we limit ourselves to the case when one of the photons is on-shell ($k^2=0$). 

Considering the  unsubtracted  dispersion relations, which result in the finite subtraction for the axial current divergence, for the cases of isovector $J^{(3)}_{\alpha 5}=  \frac{1}{\sqrt{2}}(\bar{u} \gamma_{\alpha} \gamma_5 u - \bar{d} \gamma_{\alpha} \gamma_5 d)$ and octet $J_{\alpha 5}^{(8)}=\frac{1}{\sqrt{6}}(\bar{u} \gamma_{\alpha} \gamma_5 u + \bar{d} \gamma_{\alpha} \gamma_5 d - 2 \bar{s} \gamma_{\alpha}\gamma_5 s)$ axial currents, one obtains the ASRs \cite{Horejsi:1994aj}:

\begin{equation}\label{asr}
\int_{0}^{\infty} A_{3}^{(a)}(s,q^{2}; m_i^{2}) ds =
\frac{1}{2\pi}N_c C^{(a)} \;, a=3,8,
\end{equation}
where $N_c=3$ is a number of colors, $C^{(3)}=\frac{1}{3\sqrt{2}}$ and $C^{(8)}=\frac{1}{3\sqrt{6}}$ are charge factors, $m_i$ are quark masses and $A_3$ is the imaginary part of the invariant amplitude at the tensor structure $k_{\nu} \varepsilon_{\alpha \mu \rho \sigma}
k^{\rho} q^{\sigma}$ in the variable $(k+q)^2=s>0$. The relations (\ref{asr}) are exact: $\alpha_s$ corrections are zero and it is expected that all nonperturbative corrections are absent as well (due to 't Hooft's principle \cite{Horejsi:1994aj,'tHooft:1980xb}).

As the ASRs  (\ref{asr}) do not depend on $q^2$, they remain valid also in the time-like region ($q^2>0$). The explicit way to justify the analytical continuation of ASRs  can be demonstrated by the dispersive representation for $A_3(s,q^2)$. Supposing that $A_3$ decreases fast enough at $\lvert q^2 \rvert \to \infty$ and is analytical everywhere except the cut $q^2\in(0,+\infty)$, it can be expressed as the dispersive integral without subtractions,

\begin{equation} \label{a3disp}
A_3^{(a)}(s,q^2)=\frac{1}{2\pi }\int_{0}^{\infty}dy\frac{\rho^{(a)}(s,y)}{y-q^2+i\epsilon},
\end{equation}
where  $\rho^{(a)}=2Im_{q^2} A_3^{(a)}$. Then, the ASR (\ref{asr}) for time-like $q^2$ is given by the double dispersive integral:

\begin{align} \label{asr-1}
\int_{0}^{\infty}ds\int_{0}^{\infty}dy  \frac{\rho^{(a)}(s,y)}{y-q^2+i\epsilon}=N_c C^{(a)}, \; a=3,8.
\end{align}
Note, that generally speaking, the order of integration cannot be interchanged.
The real and imaginary parts of the above ASR read:

%The integral  (\ref{a3disp}) on the line $q^2\in(0,-\infty)$ is clearly convergent, while on the line $q^2\in(0,+\infty)$ it should be defined as the principal value.
\begin{align} \label{asr-re}
p.v.\int_{0}^{\infty}ds\int_{0}^{\infty}dy \frac{\rho^{(a)}(s,y)}{y-q^2}=N_c C^{(a)},\\ \label{asr-im} \int_{0}^{\infty}ds \rho^{(a)}(s,q^2)=0,\; a=3,8.
\end{align}
%It instructive to note that from the condition (\ref{asr-im}) follows, that  $\int_{0}^{\infty}ds \rho^{(a)}(s,y)=0$.

One can easily check, that these sum rules are satisfied in the one-loop approximation for the spectral density. For a given flavor $i$ ($i=u,d,s$), it reads \cite{Horejsi:1994aj}:
\begin{align} \label{a3}
A_{3}^{(i)}=\frac{e_i^2N_c }{2\pi}\frac{\Theta (s-4m_i^2)}{(s-q^2)^2}  \left(-q^2 R^{(i)} +2m_i^2\ln\frac{1+R^{(i)}}{1-R^{(i)}}\right),
\end{align}
where $R^{(i)}=\sqrt{1-4m_i^2/s}$. %, $e_i$ are quark charges in the units of the absolute value of electron charge.
The corresponding double spectral density  in the case of massless quarks is 

\begin{align} \label{rho}
\rho^{(i)}(s,y)=e_i^2N_c y\delta'(s-y),
\end{align}
where $\delta'(x)\equiv d\delta(x)/dx$.
Note, that there are no $\alpha_s$ corrections to the one-loop expression for the spectral density (\ref{a3}) \cite{Jegerlehner:2005fs}.

\section{Transition form factors}

 As we substantiated the validity of the ASRs in the time-like region, it is useful to study their applications. The region close to $q^2=0$ is of particular interest for us, as it is hard to reach it by means of perturbative QCD while it is accessible experimentally through Dalitz decays. 

\subsection{Pion TFF in the time-like region and VMD}

Consider the\textit{ isovector channel}, i.e., the axial current is $J^{(3)}_{\alpha 5}.$ %=\frac{1}{\sqrt{2}}(\bar{u} \gamma_{\alpha} \gamma_5 u - \bar{d} \gamma_{\alpha} \gamma_5 d)$.
Saturating the lhs of the three-point correlation function (\ref{VVA}) with the resonances in the axial channel, singling out the first (pion) contribution and replacing the higher resonances' contributions with the integral of the spectral density, the ASR in the time-like region (\ref{asr-re}) leads to
\begin{equation} \label{qhd3}
\pi f_{\pi}Re F_{\pi\gamma}(q^2)+ \int_{s_3}^{\infty} A_{3}^{(3)}(s,q^{2}) ds  =\frac{1}{2\pi}N_c C^{(3)},
\end{equation}
where $s_3$ is the continuum threshold in the isovector channel, and the definitions of the meson decay constants $f_M^a$ and TFFs $F_{M\gamma}$ are as follows,

\begin{align} \label{def_f}
\langle& 0|J^{(a)}_{\alpha 5}(0) |M(p)\rangle=
i p_\alpha f^a_M, \\
\int & d^{4}x e^{ikx} \langle M(p)|T\{J_\mu (x) J_\nu(0)
\}|0\rangle=\epsilon_{\mu\nu\rho\sigma}k^\rho q^\sigma
F_{M\gamma}.
\end{align}

As the integral of $A_3$ in Eq.(\ref{def_f}) is over the region $s>s_3$, we expect that nonperturbative corrections to $A_3$ in this region are small enough and we can use the one-loop expression for it.  

Then, as $A_3^{(3)}=\frac{1}{\sqrt{2}}(A_3^{(u)}-A_3^{(d)})$, the ASR leads to the pion TFF %(for simplicity we put $m_u=m_d=m$) \cite{Klopot:2012hd},
\begin{align} 
Re F_{\pi\gamma}(q^2)=&\frac{N_c C^{(3)}}{2\pi^2f_\pi}\left[p.v.\int_{0}^{s_3}ds \int_{0}^{\infty} dy \frac{\rho^{(a)}(s,y)}{y-q^2} \right]= \nonumber \\ \label{f3m-2}
&\frac{1}{2\sqrt{2}\pi^2f_{\pi}}\frac{s_3}{s_3-q^2}.
\end{align}
%where  $R_3=\sqrt{1-4m^2/s_3}$.

We suppose, as usual, that  the continuum threshold in the axial isovector channel $s_3$ does not depend on $q^2$. The numerical value of $s_3$ was obtained in the limit $-q^2\to \infty$ of the space-like ASR \cite{Klopot:2010ke},  $s_3=4\pi^2f_\pi^2\simeq 0.67$ GeV$^2$. This expression coinsides with the one obtained  earlier from the two-point correlator analysis \cite{Radyushkin:1995pj} and is close to the numerical value obtained from two-point sum rules \cite{Shifman:1978by}.  In this way we found that the Brodsky-Lepage interpolation formula \cite{Brodsky:1981rp} (which is a one-loop approximation of the ASR) is valid also in time-like region. \footnote{The similarity between Brodsky-Lepage interpolation formula in the space-like region and the vector dominance model in the time-like region is widely known, see e.g. \cite{Behrend:1990sr}.}

We make the key observation that the TFF (\ref{f3m-2}) in the time-like region  at $q^2=s_3$ has a pole, which numerically is close to the $\rho$ meson mass squared, $m_\rho^2\simeq 0.59$ GeV$^2$. Let us stress, that in our approach $m^2_\rho=m^2_\omega=s_3$, which can be seen from the expansion of one of the electromagnetic currents in (\ref{VVA}) into isovector and isoscalar components provided that the pion continuum threshold is universal. 
This results in the relation between the axial anomaly and VMD in its simplest form when only $\rho$ (and $\omega$) contribute \footnote{The dominance of the lower-mass state(s) was pointed out also in holography approach \cite{Grigoryan:2008up} and Pad\'{e}-approximations analysis \cite{Masjuan:2012wy,Escribano:2013kba}.}.
Indeed, one can see, that the contributions of the higher mass  vector resonances ($m^2_V>s_3$)  are suppressed due to the specific form of the double spectral density (\ref{rho}). As soon as one singles out the pion contribution in the axial channel,  the photon is automatically saturated by the $\rho$ meson only, which is consistent with the VMD model. So, the specific localized form of the anomalous double spectral density (\ref{rho}) provides a foundation for VMD.

Let us note, that from (\ref{asr-im}) one can also obtain the imaginary part of the pion TFF, $Im F_{\pi\gamma}=\frac{N_c C^{a}}{2\pi f_\pi} \delta (q^2-s_3)$, which corresponds to a zero width of the $\rho$ meson. If we take into account possible small perturbative and nonperturbative corrections and/or use a more sophisticated model of continuum in the axial channel, i.e. substitute the step-function with some regularized (smoothed) one, it will lead to the finite (instead of zero) $\rho$ meson width in the vector channel. Nevertheless, this variation is significant for $Im F_{\pi\gamma}$ and $Re F_{\pi\gamma}$ only near the pole. That is why, as we are interested in the TFF far from the pole, we can neglect the imaginary part of the TFF. So, in what follows, we suppose  $ReF_{M\gamma}=F_{M\gamma}$. 

The dispersion relations for the pion TFF revealing connection to VMD model were studied before (see \cite{Landsberg:1986fd} and references therein), but, to our best knowledge without considering the connection to the axial anomaly.

Note also, that the obtained relation for the pion TFF in the time-like region $q^2>0$ (\ref{f3m-2})  naturally transforms to the expression for the pion TFF in space-like region $q^2<0$, which was obtained in \cite{Klopot:2010ke}. Therefore, Eq. (\ref{f3m-2}) gives a universal description of the pion TFF at any $q^2$.

The pole behavior (which corresponds to zero width of the $\rho$ meson) appeared since  we used the one-loop approximation for $A_3$. Therefore, Eq. (\ref{f3m-2})  can be used not too close to the pole $q^2=s_3$.  %  The  $\rho$ meson decay width can be probably reproduced by the contributions of higher corrections.

\begin{figure}
\centerline{
\includegraphics[width=0.7\textwidth]{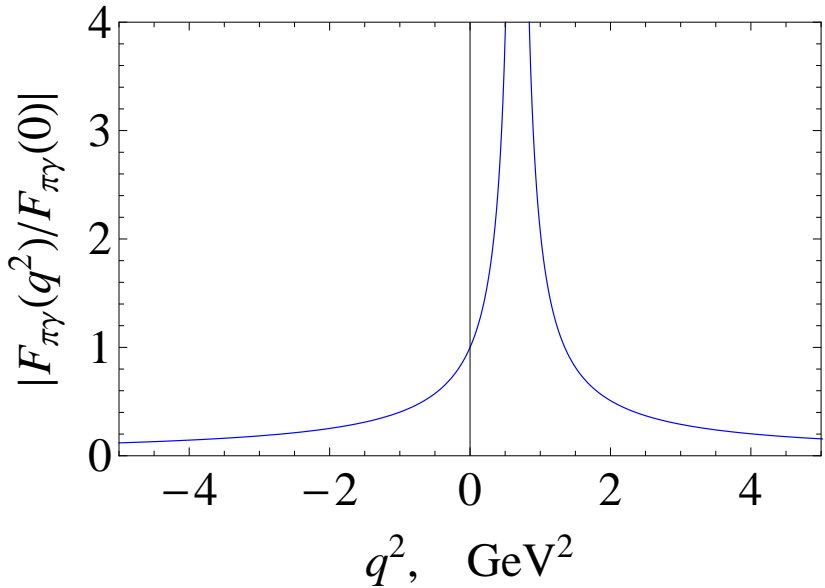}}
\caption{Pion TFF.}
\label{fig:1}
\end{figure}

The plot for the pion TFF normalized to its value at $q^2=0$, $R_\pi\equiv F_{\pi\gamma}(q^2)/F_{\pi\gamma}(0)$, is shown in Fig.\ref{fig:1}. The dimensionless slope and curvature parameters at $q^2=0$, defined  for a meson M as  $a_M=m_M^2\partial R_M/\partial q^2|_{q^2=0}$ and $b_M=\frac{1}{2}m_M^4\partial^2R_M/\partial (q^2)^2|_{q^2=0}$, are  $a_\pi= m_\pi^2/s_3 = 0.027$ and $b_\pi=m_\pi^4/s_3^2=0.73\cdot 10^{-3}$  respectively, which are in agreement with the time-like \cite{MeijerDrees:1992qb,Farzanpay:1992pz,Abouzaid:2008cd}  and space-like \cite{Behrend:1990sr,Gronberg:1997fj} experimental data extrapolations.
%The dimensionless curvature parameter $b_\pi=\frac{1}{2}m_\pi^4\partial^2R_\pi/\partial (q^2)^2|_{q^2=0}=m_\pi^4/s_3^2=0.73\cdot 10^{-3}$ is also compatible with \cite{Masjuan:2012wy}: $b_\pi = 1.06(9)_{stat}(25)_{sys}\cdot 10^{-3}$.
Our result is compatible with the ones obtained in the chiral perturbation theory (ChPT), VMD-based and some other approaches  \cite{Terschlusen:2013iqa,Landsberg:1986fd,Kampf:2005tz,Arriola:2010aq, Czyz:2012nq, Masjuan:2012wy, Dorokhov:2013xpa}. 

Our conclusions on the validity of the Brodsky-Lepage formula in the time-like region are applicable also at large $q^2$, in particular for processes $Z^0\to\pi^0\gamma$, widely discussed some time ago \cite{Deshpande:1990ej,Pham:1990mf}. Let us stress, that our previous statement (see \cite{Klopot:2012hd}, Sec. IIA) on inaccuracy of PCAC at large $q^2$ is applicable also here.

 It would be also very interesting to consider the analytical continuation of the nonperturbative correction to spectral density \cite{Klopot:2010ke,Klopot:2012hd}, so the time-like region may contribute to resolution of the controversy  between the BELLE and BABAR data.

\subsection{$\eta$ and $\eta'$ transition form factors}

Let us dwell now on the\textit{ octet channel} ($J^{(8)}_{\alpha 5}$) of the ASR. The lowest contributions to the ASR  are given by the $\eta$ and $\eta'$ mesons, both of which should be singled out due to their strong mixing. Then, employing the one-loop expression for the spectral density $A_3^{(8)}=\frac{1}{\sqrt{6}}(A_3^{(u)}+A_3^{(d)}-2A_3^{(s)})$,  taking into account $s$-quark mass contribution, but neglecting the $u,d$ quark masses, the ASR in the time-like ($q^2>0$)  region leads to 

\begin{align} \label{asr8m}
f_{\eta}^8 F_{\eta\gamma}(q^2) +f_{\eta'}^8F_{\eta'\gamma}(q^2)=\frac{1}{2\sqrt{6}\pi^2}\frac{s_8}{s_8-q^2} \Bigl [1+K_8\Bigr],
\end{align}
where the $s$-quark mass contribution $K_8$ is
\begin{align}
K_8=\frac{4m_s^2}{3s_8}(\frac{2}{R_8^{(s)}+1}+\ln\frac{1+R_8^{(s)}}{1-R_8^{(s)}}), \; R^{(s)}_8=\sqrt{1-4m_s^2/s_8}.
\end{align}
For the purposes of numerical analysis  the $s$ quark mass is  taken to be frozen to $m_s=110$ MeV at $\lvert q^2 \rvert<1$ GeV$^2$.

 The continuum threshold in the octet channel is determined \cite{Klopot:2011qq,Klopot:2012hd} from the limit $-q^2\to \infty$  of (\ref{asr8m}) and the space-like pQCD asymptotes of  $\eta, \eta'$ TFFs,

\begin{equation} \label{asr8inf}
s_8=4\pi^2((f_\eta^8)^2+(f_{\eta'}^8)^2+ 2\sqrt{2} [ f_\eta^8 f_{\eta}^0+ f_{\eta'}^8 f_{\eta'}^0]).
\end{equation}

The Eqs. (\ref{asr8m}), (\ref{asr8inf}) express the octet combination of the $\eta$ and $\eta'$ TFFs in terms of the decay constants $f_M^a$. 
For purposes of numerical analysis, we use two sets of the decay constants, obtained in the quark-flavor mixing scheme (\cite{Feldmann:1998vh} and see also \cite{Klopot:2012hd}):

 I) \cite{Klopot:2012hd} $(f_{\eta}^8,f_{\eta'}^8,f_{\eta}^0,f_{\eta'}^0)=(1.38,-0.63,0.18,1.35)f_\pi$ (in terms of quark-flavor mixing parameters $f_q=1.20f_\pi, f_s=1.65f_\pi, \phi=38.1^\circ$) and 
 
 II) \cite{Feldmann:1998vh} $(f_{\eta}^8,f_{\eta'}^8,f_{\eta}^0,f_{\eta'}^0)=(1.17,-0.46,0.19,1.15)f_\pi$ (in terms of quark-flavor mixing parameters $f_q=1.07f_\pi, f_s=1.34f_\pi, \phi=39.3^\circ$).

 The continuum threshold (\ref{asr8inf}) in terms of quark-flavor mixing scheme parameters reads: $s_8=(4/3)\pi^2(5f_q^2-2f_s^2)$. Numerically, $s_8$ for the decay constant sets I and II  is  $0.39$ GeV$^2$ and $0.48$ GeV$^2$ respectively. These numbers, similarly to the isovector channel, are close to the $\rho$ meson mass squared. Thus VMD-related argument provides an additional (to the discussed in  \cite{Klopot:2011ai}) explanation  why the continuum threshold $s_8$ is so surprisingly small. Indeed, the similarity of duality intervals in the isovector and octet channels and respective positions of poles in the time-like regions is required for consistency of the anomaly-motivated VMD explanation.

At the same time, the numerical value of the continuum threshold in the octet channel is determined worse than  the one in the isovector channel because of the uncertainty in the decay constants obtained in different analyses. As the ASR can be reliably used in the region sufficiently far away from the pole, this discrepancy is not very important. 

\begin{figure}
\centerline{
\includegraphics[width=0.7\textwidth]{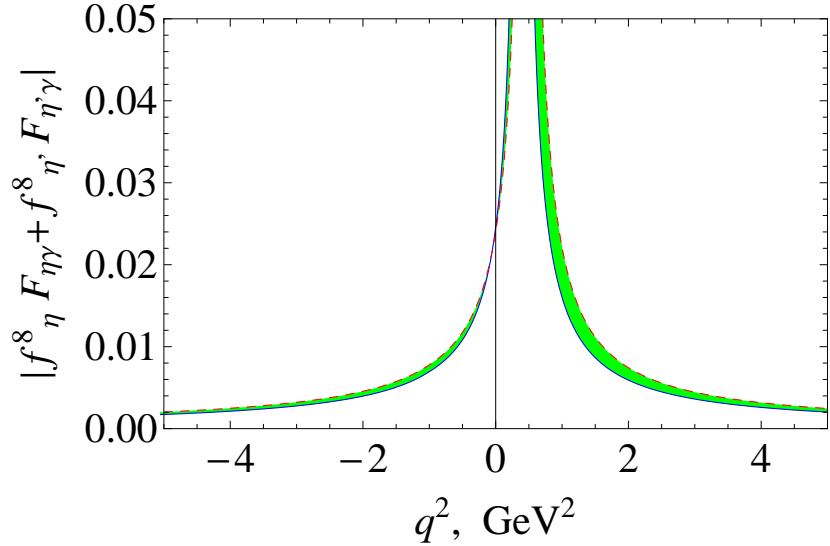}}
\caption{The octet combination of the TFFs (\ref{asr8m}). % (i.e. lhs of the Eq. (\ref{asr8m})). % (evolved from $m_s=95$ MeV at $Q^2=2$  GeV$^2$ [PDG2012]).
}
\label{fig:2}
\end{figure}

From Fig.\ref{fig:2} the reliable region of  applicability of Eq. (\ref{asr8m}) can be estimated as $q^2\lesssim 0.2$ GeV$^2$ and  $q^2\gtrsim 2$ GeV$^2$. The band indicates the region between the lines corresponding to the considered sets (I and II) of decay constants.

In order to obtain the separate expressions for the $\eta$ and $\eta'$ TFFs, let us bring another relation \cite{Klopot:2012hd}. Basing on the widely used hypothesis (see e.g. \cite{BABAR:2011ad}), that the TFFs of the (unphysical) state  $|n \rangle \equiv \frac{1}{\sqrt{2}}(|\bar{u}u\rangle+|\bar{d}d\rangle)$ is related to the pion TFF as $F_{n\gamma}(q^2)=(5/3) F_{\pi\gamma}(q^2)$ (the numerical factor originates from the quark charges $(e_u^2+e_d^2)/(e_u^2-e_d^2)=5/3$), one obtains $\frac{5}{3}F_{\pi\gamma}=F_{\eta\gamma} \cos \phi + F_{\eta'\gamma} \sin \phi$ (quark-flavor mixing scheme is implied). Combining this equation with (\ref{asr8m}), (\ref{asr8inf}) and  (\ref{f3m-2}),  we obtain the  $\eta$ and $\eta'$ TFFs:

%\begin{equation} \label{Fq}
%\frac{5}{3}F_{\pi\gamma}=F_{\eta\gamma} \cos \phi + F_{\eta'\gamma} \sin \phi.
%\end{equation}

%Now, combining (\ref{asr8m}), (\ref{asr8inf}) and  (\ref{f3m-2}), (\ref{Fq}), we obtain the  $\eta$ and $\eta'$ TFFs:

\begin{align}
F_{\eta\gamma}(q^2)&=\frac{5}{12\pi^2f_s f_\pi}\frac{s_3(\sqrt{2}f_s\cos \phi -f_q\sin \phi)}{s_3-q^2} + 
\frac{1}{4\pi^2f_s} \frac{s_8\sin\phi}{s_8-q^2}(1+K_8), \label{feta}  \\
F_{\eta'\gamma}(q^2)&=\frac{5}{12\pi^2f_sf_\pi} \frac{s_3(\sqrt{2}f_s\sin \phi +f_q\cos \phi)}{s_3-q^2} - 
\frac{1}{4\pi^2f_s} \frac{s_8\cos\phi}{s_8-q^2}(1+K_8), \label{fetap}
\end{align}
%where $s_3=4\pi^2f_\pi^2$, $s_8=(4/3)\pi^2(5f_q^2-2f_s^2)$. 

\begin{figure}
\centerline{
\includegraphics[width=0.7\textwidth]{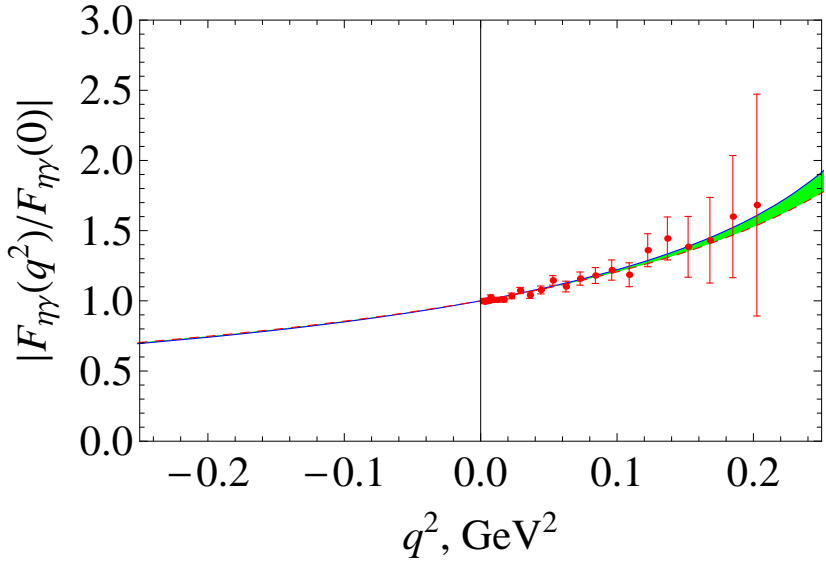}}
\caption{The $\eta$ meson TFF (\ref{feta}) and A2 Collaboration data \cite{Aguar-Bartolome:2013vpw}.}
\label{fig:3}
\end{figure}

In  Fig. \ref{fig:3} the plot of the $\eta$ TFFs (\ref{feta}) (normalized to its value at $q^2=0$) is given in the range of $q^2\in(-0.25, 0.25)$ GeV$^2$. 
The blue solid and red dashed curves correspond to the parameter sets I and II respectively, the green band indicates the region between the curves. The data from A2 Collaboration  %(Crystal Ball/TAPS detector at MAMI)
\cite{Aguar-Bartolome:2013vpw} is presented by the red points with error bars. Both decay constants sets (I and II) give  $\chi^2/24=0.11$.
We see a good description of the experimental data. The slope and curvature parameters of the line at $q^2=0$ are $a_{\eta}=0.54, b_{\eta}=0.31$ ($a_{\eta}=0.51, b_{\eta}=0.27$) for the decay constant set I (II). These numbers are close to the results of  \cite{Escribano:2013kba,Hanhart:2013vba}, ChPT \cite{Ametller:1991jv}, time-like \cite{Arnaldi:2009aa} and space-like \cite{Behrend:1990sr,Gronberg:1997fj} experimental data fits.

The data of NA60 experiment for the $\eta$ TFF \cite{Arnaldi:2009aa}, available in the region of $q^2\in(0.06, 0.21)$ GeV$^2$ (not shown in Fig. \ref{fig:3}), is also in a good agreement with our result (\ref{fetap}). 
Let also note, that the result of NA60 for the $\omega-\pi^0$ time-like TFF (in the process $\omega\to\mu^+\mu^-\pi^0$), where the discrepancy with the VMD model was found, is related to the axial anomaly with two virtual photons, which is beyond the scope of this paper.

\begin{figure}
\centerline{
\includegraphics[width=0.7\textwidth]{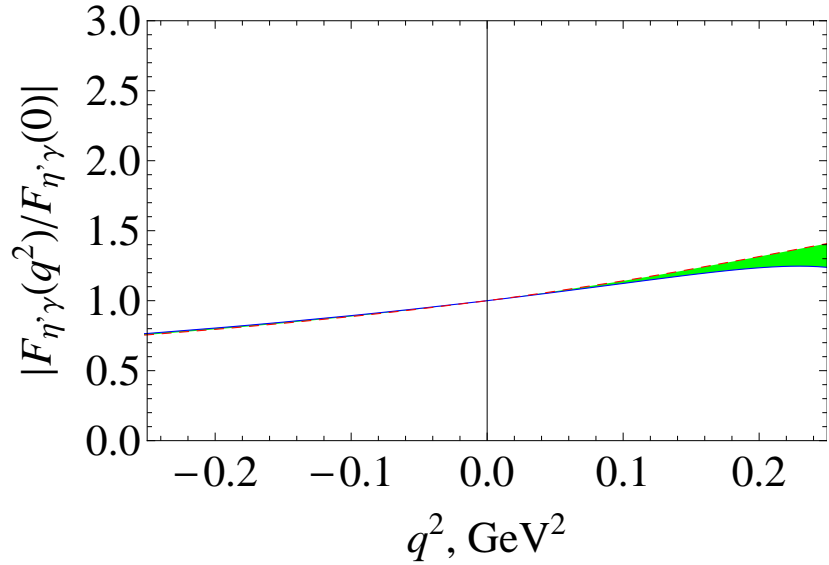}}
\caption{The $\eta'$ meson TFF (\ref{fetap}).}
\label{fig:4}
\end{figure}

In  Fig.\ref{fig:4} the plot of the $\eta'$ TFF (\ref{fetap}) (normalized to $F_{\eta'\gamma}(0)$) is given in the range of $q^2\in(-0.25, 0.25)$ GeV$^2$. 
The blue solid and red dashed curves correspond to the parameter sets I and II respectively. The green band indicates the region between the curves.
The slope and curvature parameters of the line at $q^2=0$ are $a_{\eta}=1.06, b_{\eta}=0.76$ ($a_{\eta}=1.16, b_{\eta}=1.19$) for the decay constant set I (II). These numbers are may be compared with \cite{Escribano:2013kba}: $a_{\eta'} = 1.37(18)$, $b_{\eta'} = 1.94(67)$.

In summary, we have done the analytical continuation of the ASRs to the time-like region, and obtained the TFFs of $\pi^0$, $\eta$ and $\eta'$ mesons in this region. The excellent agreement with the available experimental data is found. So, we completed our former works \cite{Klopot:2010ke,Klopot:2011qq,Klopot:2011ai,Klopot:2012hd} and now describe the TFFs in both time-like and space-like regions. Based on the ASR, we have substantiated the VMD model for the TTFs  with one real and one virtual photon.

We would like to thank M.~Amaryan, E.~Kuraev, S.~Mikhailov and N.~Stefanis for stimulating discussions and comments. This work is supported in part by RFBR, research projects 12-02-00613, 12-02-91526, 12-02-00284, 13-02-01060 and Heisenberg-Landau Program (JINR).

%\begin{comment}

%\end{comment}

\end{document}